\documentclass{Interspeech2024}
\usepackage{booktabs}
\usepackage{graphicx}
\usepackage{amsmath}
\usepackage{multirow}
\usepackage{tabularx}
\usepackage{array}
\usepackage{eso-pic}
\usepackage{graphicx} 
\usepackage{hyperref}
\usepackage[capitalize]{cleveref} 
\hypersetup{
    colorlinks,
    linkcolor={red!50!black},
    citecolor={blue!50!black},
    urlcolor={blue!80!black}
}

\usepackage[draft]{pdfcomment} 
\usepackage[hyperfirst=false,acronym,sort=none,shortcuts,nopostdot,style=super,nonumberlist,toc,nogroupskip]{glossaries}
\usepackage{xcolor} 

\usepackage{xspace}

\glsdisablehyper 

\definecolor{Sepia}{RGB}{112,66,20}

\definecolor{AcColor}{rgb}{0,0,0}

\newcommand{\accolor}[1]{\textcolor{AcColor}{#1}} 

\newacronymstyle{myacro}
{%
  \GlsUseAcrEntryDispStyle{long-short}%
}%
{%
  \GlsUseAcrStyleDefs{long-short}%
}

\setacronymstyle{myacro}

\let\xx\tip
\let\xxx\tips

\newacronym{adc}{ADC}{Analog to Digital Converter}
\newacronym{admm}{ADMM}{Alternating Direction Method of Multipliers}
\newacronym{asic}{ASIC}{Application Specific Integrated Circuit}
\newacronym{cem}{CEM}{Cross-Entropy Method}
\newacronym{cots}{COTS}{Commodity Off-The-Shelf}
\newacronym{cpu}{CPU}{Central Processing Unit}
\newacronym{cnn}{CNN}{Convolutional Neural Network}
\newacronym{ddp}{DDP}{Differential Dynamic Programming}
\newacronym{dgrnn}{DG-RNN}{Dynamic Gated Recurrent Neural Network}
\newacronym{dgru}{D-GRU}{Dynamic Gated Recurrent Unit}
\newacronym{dnn}{DNN}{Deep Neural Network}
\newacronym{dns}{DNS}{Deep Noise Suppression}
\newacronym{dof}{DOF}{Degree Of Freedom}
\newacronym{dram}{DRAM}{Dynamic RAM}
\newacronym{fc}{FC}{Fully Connected}
\newacronym{fpga}{FPGA}{Field Programmable Gate Array}
\newacronym{gpu}{GPU}{Graphics Processing Unit}
\newacronym{gru}{GRU}{Gated Recurrent Unit}
\newacronym{hil}{HIL}{Hardware In the Loop}
\newacronym{ip}{IP}{Interior Point}
\newacronym{ipb}{IP block}{Intellectual Property Block}
\newacronym{lstm}{LSTM}{Long Short Term Memory}
\newacronym{ltv}{LTV}{Linear Time Varying}
\newacronym{mac}{MAC}{Multiply-Accumulate}
\newacronym{mae}{MAE}{Median Angular Error}
\newacronym{mlp}{MLP}{Multilayer Perceptron}
\newacronym{mee}{MEE}{Median Endpoint Error}
\newacronym{mpc}{MPC}{Model Predictive Control}
\newacronym{mpcer}{MPC}{Model Predictive Controller}
\newacronym{mppi}{MPPI}{Model Predictive Path Integral}
\newacronym{nc}{NC}{Neural Controller}
\newacronym{nmpc}{NMPC}{Nonlinear Model Predictive Control}
\newacronym{nn}{NN}{Neural Network}
\newacronym{npc}{NPC}{Neural Predictive Control}
\newacronym{npu}{NPU}{Neural Processing Unit}
\newacronym{ode}{ODE}{Ordinary Differential Equation}
\newacronym{pcb}{PCB}{Printed Circuit Board}
\newacronym{pd}{PD}{Proportional Derivative}
\newacronym{pid}{PID}{Proportional Integral Derivative}
\newacronym{pl}{PL}{Programmable Logic}
\newacronym{ps}{PS}{Processing System}
\newacronym{pso}{PSO}{Particle Swarm Optimization}
\newacronym{pwm}{PWM}{Pulse Width Modulation}
\newacronym{qp}{QP}{Quadratic Programming}
\newacronym{rl}{RL}{Reinforcement Learning}
\newacronym{rnn}{RNN}{Recurrent Neural Network}
\newacronym{ros}{ROS}{Robot Operating System}
\newacronym{rpgd}{RPGD}{Resampling Parallel Gradient Descent}
\newacronym{slp}{SLP}{Single Layer Perceptron}
\newacronym{sm}{SM}{Supplementary Material}
\newacronym{snr}{SNR}{Signal-to-Noise Ratio}

\newacronym[description={System on Chip; FPGA with embedded programmable processor}]{soc}{SoC}{System on Chip}
\newacronym{sqp}{SQP}{Sequential Quadratic Programming}
\newacronym{sram}{SRAM}{Static RAM}
\newacronym{usb}{USB}{Universal Serial Bus}
\newacronym{vga}{VGA}{Video Graphics Adaptor}
\newacronym{uart}{UART}{Universal Asynchronous Receiver/Transmitter}
\newacronym{xla}{XLA}{Accelerated Linear Algebra}

\newacronym{mpt}{MPT}{Multi-Path Transformer}
\newacronym{dprnn}{DPRNN}{Dual-Path RNN}

\usepackage{lipsum}

\newcommand\blfootnote[1]{%
  \begingroup
  \renewcommand\thefootnote{}\footnote{#1}%
  \addtocounter{footnote}{-1}%
  \endgroup
}

\interspeechcameraready

\title{Dynamic Gated Recurrent Neural Network for\\Compute-efficient Speech Enhancement}

\name[affiliation={1}]{Longbiao}{Cheng}
\name[affiliation={2}]{Ashutosh}{Pandey}
\name[affiliation={2}]{Buye}{Xu}
\name[affiliation={1}]{Tobi}{Delbruck}
\name[affiliation={1}]{Shih-Chii}{Liu}

\address{
   $^1$ Institute of Neuroinformatics, University of Zurich and ETH Zurich \\
   $^2$ Reality Labs Research, Meta
  }
\email{\{longbiao, tobi, shih\}@ini.uzh.ch, \{apandey620, xub\}@meta.com}


\begin{document}

\maketitle

\begin{abstract}
This paper introduces a new Dynamic Gated Recurrent Neural Network (DG-RNN) for compute-efficient speech enhancement models running on resource-constrained hardware platforms.
It leverages the slow evolution characteristic of RNN hidden states over steps, and updates only a selected set of neurons at each step by adding a newly proposed select gate to the RNN model. This select gate allows the computation cost of the conventional RNN to be reduced during network inference.
As a realization of the DG-RNN, we further propose the Dynamic Gated Recurrent Unit (D-GRU) which does not require additional parameters. 
Test results obtained from several state-of-the-art compute-efficient RNN-based speech enhancement architectures using the DNS challenge dataset, 
show that the D-GRU based model variants maintain similar speech intelligibility and quality metrics comparable to the baseline GRU based models even with an average 50\% reduction in GRU computes.

\blfootnote{This work was supported by a research contract from Meta Reality Labs Research.}

\end{abstract}

\section{Introduction}
The presence of ambient noise in real-world environments can diminish significantly the speech quality and intelligibility of a speaker. This degradation adversely affects the listener's experience in various applications involving telecommunication systems, online meeting platforms, and Augmented Reality (AR) / Mixed Reality (MR) devices. Despite the promising advancements made in deep-learning-based speech enhancement algorithms~\cite{wang2018supervised}, their deployment on resource-constrained platforms and subsequent run-time latency is impacted by incompatible memory and compute requirements needed of these platforms.

U-Net based architectures are one class of architectures used extensively for speech enhancement~\cite{tan2018convolutional, tan2019complex, hu2020dccrn, li2021two, le2021dpcrn, defossez2020real}. At its core, the model typically features a \xx{cnn}-based encoder and decoder along with a temporal information modeling layer. The U-Net architecture delivers great results in reducing noise while preserving the quality and intelligibility of the original speech, primarily because of the skip connections between the encoder and decoder layers. However these connections require that the intermediate feature maps be stored therefore requiring memory that is not easily available on mobile platforms.

\begin{figure*}[th]
  \centering
  \includegraphics[width=0.85\linewidth]{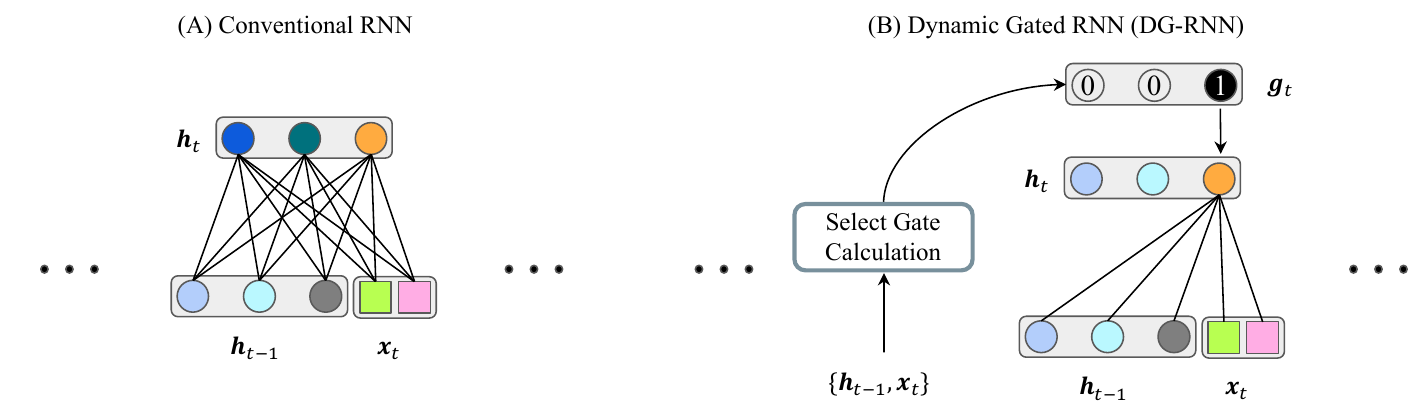}
  \vspace*{-2.5mm}
  \caption{
  Illustration of the update processes of {\rm (A)} conventional Recurrent Neural Network (RNN) and {\rm (B)} Dynamic Gated Recurrent Neural Network (DG-RNN) at step $t$. {\rm (A)} For conventional RNN, all neurons in the hidden state are updated at each step. {\rm (B)} DG-RNN first identifies which neurons need updating, as indicated by a $1$ in the proposed select gate $ \boldsymbol{g}_t $. When a neuron is marked with $1$, it undergoes the RNN update process. Those marked with $0$ retain their values from the previous hidden state. 
  }
  
  \label{fig:D_RNN}
  \vspace*{-5mm}

\end{figure*}

Recent studies show that \xx{rnn}-based architectures can match the speech enhancement performance of the U-Net architectures while using less hardware resource. For instance, FullSubNet~\cite{hao2021fullsubnet,chen2022fullsubnet+} employs a two-layer \xx{rnn} architecture to first model the full-band global information, which is then used in conjunction with another two layers of \xxx{rnn} to estimate the masks at each time-frequency bin. Similarly, TaErLite~\cite{li2023general} segments the enhancement process into two phases, first, coarse magnitude spectrum enhancement and, then, multi-stage complex spectrum refinement, utilizing \xxx{rnn} as the core component in both phases. The introduction of dual-path~\cite{luo2020dual, dang2022dpt} and multi-path~\cite{pandey2022tparn} strategies in \xx{rnn}-based models significantly reduce parameter size by reusing smaller \xxx{rnn} across various dimensions, such as different frames or frequency bins. Additional methods that incorporate band splitting~\cite{yu2022high, chen2023complexity} and frame skipping~\cite{chen2023ultra} strategies further reduce the computation costs of these models.

This work proposes a new Dynamic Gated Recurrent Neural Network (DG-RNN) framework that can 
decrease further the computation cost of these \xx{rnn}-based models during inference. The framework was tested on a network trained for a regression task, in this case, speech enhancement. The contributions of the work are as follows: 
\begin{itemize}

\item 
A novel \xx{rnn} architecture, called \xx{dgrnn}, which updates only a selected subset of neurons at each step. \xx{dgrnn} incorporates a new select gate to the conventional \xx{rnn} for determining the dynamic set of neurons to be updated at each step. Neurons that are not selected will skip their update process at that step, resulting in a computation reduction of the conventional \xx{rnn}.

\item 
The application of \xx{dgrnn}'s selective update strategy to the conventional \xx{gru},  leading to the creation of \xx{dgru}. Notably, the \xx{dgru} utilizes the update gate of conventional \xx{gru} to derive the select gate, without the introduction of additional parameters.

\item Comprehensive experimental validation of \xx{dgru} in a single-channel speech enhancement task using the \xx{dns} challenge dataset~\cite{reddy2020interspeech}. Our evaluation encompasses models that solely employ a two-layer \xx{gru}, as well as models that use \xx{mpt} blocks~\cite{chen2023complexity} or \xx{dprnn} blocks~\cite{luo2020dual}. Results show minimal impact on speech quality and intelligibility with a 50\% reduction in GRU computation costs even under -5 dB \xx{snr} conditions. Audio samples can be found \href{https://chenglongbiao.github.io/DG-RNN/}{here}. 

\end{itemize}

\section{Dynamic Gated RNN (DG-RNN)}

As shown in previous studies~\cite{neil2017delta}, natural inputs to a recurrent neural network tend to have a high degree of temporal auto-correlation, resulting in slowly changing  hidden states. From these observations, we propose the \xx{dgrnn} framework, designed to reduce the computation cost of conventional \xxx{rnn}. In this framework, only a percentage of neurons is updated at each step, while the remaining neurons retain their previous state value, based on the assumption that their hidden state value underwent minimal change during the current step. 

The selective update of the proposed \xx{dgrnn} is depicted in Fig.~\ref{fig:D_RNN}.
Upon receiving an input $ \boldsymbol{x}_t $ at step $ t $, the \xx{dgrnn} first determines the set of neurons that will be updated at the current step using our new proposed select gate, $\boldsymbol{g}_t$. The select gate is computed using $ \boldsymbol{x}_t $ and the previous hidden state $ \boldsymbol{h}_{t-1} $:
\begin{equation} 
\boldsymbol{g}_t = \mathcal{B} (\mathcal{G}(\boldsymbol{x}_t, \boldsymbol{h}_{t-1})), \label{eq:DGRNN_g} 
\end{equation} 
where $ \mathcal{G}(\cdot) $ represents the computation process of a neural network, and $ \mathcal{B}(\cdot) $ denotes a binarization operation, yielding $ \boldsymbol{g}_t $ as a binary vector of length $J$, where $J$ is the number of total neurons.
Based on the output of the select gate, the hidden state in \xx{dgrnn} at step $ t $ is updated as follows:
\begin{equation}
    h_t^j =
    \begin{cases} 
    \mathcal{F}^j (\boldsymbol{x}_t, \boldsymbol{h}_{t-1}) & \text{if } g_t^j = 1,  \\
    h_{t-1}^j & \text{if } g_t^j = 0. \label{eq:DGRNN_h}
    \end{cases}
\end{equation}
where $h_t^j$ indicates the state of the $j$-th neuron at step $t$ and $\mathcal{F}^j(\cdot)$ is its update function.

We additionally define the \textbf{neuron update percentage $\mathcal{P}_t$} at step $t$ as $\mathcal{P}_t = (\sum_{j=1}^J g_t^j)/J * 100$. Note that $\mathcal{P}$ is the most important parameter in determining cost savings; reducing $\mathcal{P}$ as much as possible without compromising accuracy will lead to lower costs. Moreover, the select gate generation process should be lightweight to not negate the advantages of sparse update. In Section~\ref{sec:dGRU}, we demonstrate that by applying \xx{dgrnn}'s selective update strategy to the \xx{gru}, the select gate can be derived without additional parameters.

\section{Dynamic GRU (D-GRU)}
\label{sec:dGRU}

This section begins with a brief summary of the neuron update process in a conventional \xx{gru}, followed by the introduction of the \xx{dgru}.
Then cost savings of the \xx{dgru} compared to the conventional \xx{gru} are presented.

\subsection{\xx{gru} Update}
The update equations for a conventional \xx{gru} are:
\begin{align}
\boldsymbol{r}_t &= \sigma(\mathbf{W}_{ir}\boldsymbol{x}_t + \boldsymbol{b}_{ir} + \mathbf{W}_{hr}\boldsymbol{h}_{(t-1)} + \boldsymbol{b}_{hr}) \\
\boldsymbol{z}_t &= \sigma(\mathbf{W}_{iz}\boldsymbol{x}_t + \boldsymbol{b}_{iz} + \mathbf{W}_{hz}\boldsymbol{h}_{(t-1)} + \boldsymbol{b}_{hz}) \label{eq:gru_z} \\
\boldsymbol{c}_t &= \tanh(\mathbf{W}_{ic}\boldsymbol{x}_t + \boldsymbol{b}_{ic} + \boldsymbol{r}_t \ast (\mathbf{W}_{hc}\boldsymbol{h}_{t-1} + \boldsymbol{b}_{hc})) \\
\boldsymbol{h}_t &= \boldsymbol{z}_t \ast \boldsymbol{c}_t + (1 - \boldsymbol{z}_t) \ast \boldsymbol{h}_{t-1} \label{eq:gru_h}
\end{align}
where $\bm{h}_t$ is the hidden state at step $t$, $\bm{x}_t$ is the input, and $\bm{r}_t$, $\bm{z}_t$, $\bm{c}_t$ are the reset, update gates and candidate hidden state, respectively. $\sigma$ is the sigmoid function and $\ast$ is the Hadamard product.

\subsection{\xx{dgru} Update}
As shown in Eq.~(\ref{eq:gru_h}),  the new hidden state $\bm{h}_t$ is the weighted average of the previous hidden state $\bm{h}_{t-1}$ and the current candidate  state $\bm{c}_t$. This process is determined by the update gate $\bm{z}_t$, in which every element $z^j_t$ has a value between 0 and 1 according to Eq.~(\ref{eq:gru_z}).  
When $z^j_t$ is close to 1, the hidden state $h^j_t$ is largely replaced by the candidate hidden state $c^j_t$. Conversely, $z^j_t$ close to 0 means that $h^j_{t}$ is close to $h^j_{t-1}$.

In our proposed \xx{dgru}, we only update neurons with the top-$A$ largest values in $\bm{z}_t$, i.e., the select gate output is mathematically represented as:
\begin{equation}
    g^j_t = 
    \begin{cases} 
    1 & \text{if } z^j_t \text{ is among the top-}A \text{ largest elements of } \bm{z}_t \\
    0 & \text{otherwise} \label{eq:DGRU_g}
    \end{cases}
\end{equation}
where $A=(\mathcal{P}/100)*J$ is the number of neurons that need to be updated, with $\mathcal{P}$ being the predefined constant update percentage across steps. 

Note that instead of selecting neurons based on the top-$A$ largest $z^j_t$ values, we could also select neurons based on a set  threshold,  $\theta^j$:
\begin{equation}
\label{eq:delta_g}
    g^j_t = 
    \begin{cases} 
    1 & \text{if } z^j_t > \theta^j \\
    0 & \text{if } z^j_t \leq \theta^j
    \end{cases}.
\end{equation}
However, this approach results in varying numbers of neurons that require updating in each step, leading to dynamically changing computation loads per step. For practical implementations, the hardware still needs to be capable of doing all the compute for updating of all neurons in order to satisfy the real-time processing constraint. The performance of the threshold based select gate will be discussed in future work.

\begin{table*}[]
\caption{Parameter sizes and computation costs of tested models and objective results of them on DNS testset; see \cref{sec:results}.}
\vspace*{-2.5mm}
\label{tab:all_results}
\centering

\resizebox{\linewidth}{!}{%

\begin{tabular}{@{}ccccccrcccc@{}}
\toprule
\multirow{2}{*}{\small{Model}} & \multirow{2}{*}{\small{Para. (M)}} & \multirow{2}{*}{\small{{GRU type}}} & \multirow{2}{*}{\small{$\mathcal{P}$ (\%)}} & \multicolumn{3}{c}{\small{\xx{mac} (M/s)}}                                     & \multirow{2}{*}{\small{PESQ}} & \multirow{2}{*}{ESTOI} & \multirow{2}{*}{\small{SISNR}}  & \multirow{2}{*}{\small{DNS$_{\rm OVAL}$}} \\ \cmidrule(lr){5-7}
                   &                              &                           &                                 & \small{non-GRU}                 & \small{GRU}     & \multicolumn{1}{c}{\small{All layers}} &                       &                       &                        &                        \\ \midrule \midrule
noisy                           & -                     & -                          & -             & -                       & -       & \multicolumn{1}{c}{-}          & 1.58                  & 81          & 9.23           & 2.48          \\ \midrule
\multirow{4}{*}{\textit{GRU}}   & \multirow{4}{*}{1.34} & \xx{gru}                   & 100           & \multirow{4}{*}{10.34}  & 124.98  & 135.32 (100\%)                  & 2.58                  & \textbf{90} & 15.47          & \textbf{3.22} \\ \cline{3-3}
                                &                       & \multirow{3}{*}{\xx{dgru}} & 75            &                         & 104.15  & 114.48 \ \ (84\%)                   & \textbf{2.62}         & \textbf{90} & \textbf{15.68} & \textbf{3.22} \\
                                &                       &                            & 50            &                         & 83.31   & 93.65  \ \ (69\%)                    & 2.58                  & \textbf{90} & 15.59          & 3.20          \\
                                &                       &                            & 25            &                         & 62.49   & 72.83  \ \ (53\%)                    & 2.45                  & 89          & 14.82          & 3.15          \\ \midrule
\multirow{4}{*}{\textit{MPT}}   & \multirow{4}{*}{0.28} & \xx{gru}                   & 100           & \multirow{4}{*}{71.61}  & 210.71  & 282.32 (100\%)                  & \textbf{2.66}         & \textbf{90} & \textbf{15.04} & \textbf{3.25} \\ \cline{3-3}
                                &                       & \multirow{3}{*}{\xx{dgru}} & 75            &                         & 175.58  & 247.19  \ \ (87\%)                   & 2.65                  & \textbf{90} & 14.96          & 3.23          \\
                                &                       &                            & 50            &                         & 140.46  & 212.06  \ \ (75\%)                   & 2.63                  & \textbf{90} & 14.96          & 3.22          \\
                                &                       &                            & 25            &                         & 105.36  & 176.96  \ \ (62\%)                   & 2.57                  & \textbf{90} & 14.75          & 3.20          \\ \midrule
\multirow{4}{*}{\textit{DPRNN}} & \multirow{4}{*}{0.35} & \xx{gru}                   & 100           & \multirow{4}{*}{811.02} & 4878.35 & 5689.37 (100\%)                 & 2.85                  & \textbf{92} & \textbf{18.08} & \textbf{3.29} \\ \cline{3-3}
                                &                       & \multirow{3}{*}{\xx{dgru}} & 75            &                         & 4065.13 & 4876.14  \ \ (85\%)                  & \textbf{2.88}         & \textbf{92} & 17.71          & 3.26          \\
                                &                       &                            & 50            &                         & 3251.91 & 4062.92  \ \ (71\%)                  & 2.82                  & \textbf{92} & 17.74          & 3.27          \\
                                &                       &                            & 25            &                         & 2439.18 & 3250.19  \ \ (57\%)                  & 2.72                  & 91          & 17.28          & 3.24          \\ \bottomrule
\end{tabular}
}
\vspace*{-5mm}
\end{table*}

Following Eq.~(\ref{eq:DGRNN_h}) and Eq.~(\ref{eq:gru_h}), the hidden state of a neuron in the \xx{dgru} is updated as follows:
\begin{equation}
h^j_t = 
\begin{cases} 
z^j_t \ast c_t^j + (1 - z^j_t) \ast h^j_{t-1} & \text{if } g^j_t = 1 \\
h^j_{t-1} & \text{if } g^j_t = 0 \label{eq:DGRU_h}
\end{cases}.
\end{equation}
The computations for $r^j_t$ and $c^j_t$ hence can be skipped for neurons whose $g^j_t = 0$: 
\begin{align}
r^j_t &= 
    \begin{cases} 
    \sigma(\bm{w}^j_{ir}\bm{x}_t + b^j_{ir} + \bm{w}^j_{hr}\bm{h}_{t-1} + b^j_{hr}) & \text{if } g^j_t = 1 \\
    \text{(skipped)} & \text{if } g^j_t = 0
    \end{cases} \label{eq:DGRU_r} \\
c^j_t &=
    \begin{cases} 
    \tanh \left( \bm{w}^j_{ic}\bm{x}_t + b^j_{ic} + r^j_t \left( \bm{w}^j_{hc}\bm{h}_{t-1} + b^j_{hc} \right) \right) & \text{if } g^j_t = 1 \\
    \text{(skipped)} & \text{if } g^j_t = 0
    \end{cases} \label{eq:DGRU_c}.
\end{align}

\subsection{Cost Savings}
\label{sec:cost_saving}

There is an additional cost that is incurred for determining the neurons to update following Eq.~(\ref{eq:DGRU_g}). 
However, the computational complexity of this process is $\mathcal{O}(J)$~\cite{cormen2022introduction} and is negligible compared to other costs of updating the \xx{gru} equations.
Following Eqs.~(\ref{eq:DGRU_r}, \ref{eq:DGRU_c}), we require $\mathcal{P}$ percent of the original computation costs associated with the reset gates and candidate hidden states. Since the computation for the update gate $\bm{z}$ cannot be saved, the total computation of the \xx{dgru} is $(1+2\mathcal{P}/100)/3$ of that in the conventional \xx{gru}.

\section{Experimental Setup}
\begin{figure*}[th]
  \centering
    
  \includegraphics[width=\linewidth]{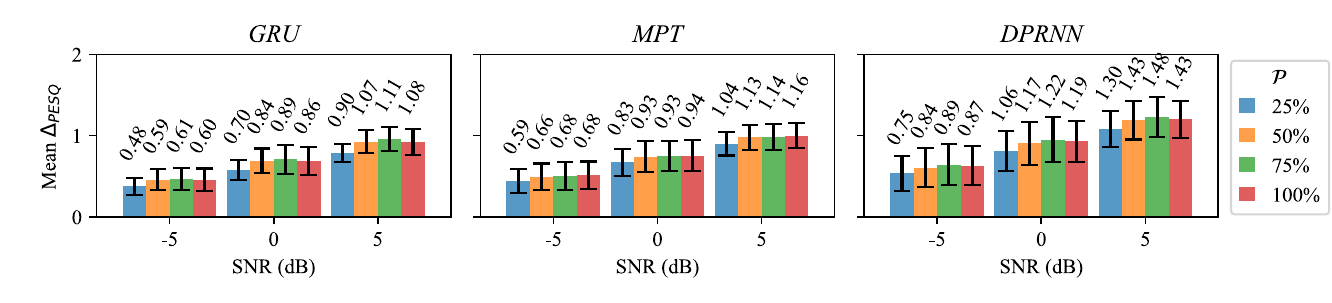}
  \vspace*{-9mm}
  \caption{Mean PESQ improvement ($\Delta_{PESQ}$) of different update percentages $\mathcal{P}$ on different models under \{-5, 0, 5\} dB SNRs. }
  \label{fig:SNR}
  \vspace*{-2.5mm}

\end{figure*}

The \xx{dgru} is tested on a set of speech enhancement models described in Section~\ref{sec:dGRU_SE} while the dataset and tested \xx{snr} ranges are described in Section~\ref{sec:dataset_SE}.

\subsection{\xx{dgru} Based Speech Enhancement}
\label{sec:dGRU_SE}

We applied \xx{dgru} on three strong causal speech enhancement models that here we call 1) \textit{GRU}~\cite{narayanan2013ideal}; 2) \textit{\xx{mpt}}~\cite{chen2023complexity}; and 3) the time-frequency domain \textit{\xx{dprnn}}~\cite{luo2020dual}.
\begin{itemize}
    \item The \textit{GRU} model comprises a \xx{fc} layer, two \xx{gru} layers, and another \xx{fc} layer. The output dimensions of the first \xx{fc} layer and the two \xx{gru} layers are 320. The final \xx{fc} layer outputs the ideal ratio mask (IRM)~\cite{narayanan2013ideal}. 
    \item The \textit{\xx{mpt}} model~\cite{chen2023complexity} consists of 2 \xx{mpt} blocks, each containing four \xxx{gru}, with input and output dimensions of 24 and 48, respectively. Following the \xx{mpt} paper, this model outputs both the IRM and magnitude deep-filtering coefficients~\cite{schroter2022deepfilternet}. 
    \item The time-frequency domain \textit{\xx{dprnn}}~\cite{luo2020dual} directly outputs the estimated clean speech’s complex spectrogram. It includes four dual-path blocks, each composed of three \xxx{gru}, where both input and output dimensions of each \xx{gru} are 64. 
\end{itemize}
For the \textit{GRU} and \textit{MPT} models which enhance only the magnitude, the input is the noisy magnitude spectrum, and the loss functions are the mean square error (MSE) between the enhanced and target magnitude spectrograms. For the \textit{\xx{dprnn}}, with a complex spectrogram as input, the training loss function is an average of the magnitude MSE and the complex spectrogram MSE. For each model, the neuron update percentage $\mathcal{P}$ is set to [100, 75, 50, 25] and the effective computation cost for the \xx{dgru} at each $\mathcal{P}$ are [100\%, 83.33\%, 66.66\%, 50\%], respectively (see Section ~\ref{sec:cost_saving}). The parameter size and the number of \xx{mac} operations, with a breakdown between the non-\xx{gru} and \xx{gru} layers costs, for all three types of models and different $\mathcal{P}$ values are detailed in Table~\ref{tab:all_results}.

We highlight that each of the three model types is characterized by their lightweight design, marked by either parameter efficiency, low computation cost, or a combination of both. These characteristics pose a challenge in demonstrating the benefits of the \xx{dgru}, given that these models are already designed for minimal computation cost and size compactness. 

\subsection{Dataset}
\label{sec:dataset_SE}

The speech enhancement models are trained on the \xx{dns} challenge dataset~\cite{reddy2020interspeech}, a benchmark dataset  used for assessment of speech enhancement algorithms.  
The dataset comprises 500 hours of clean speech from 2150 speakers and 180 hours of noise signals. Noisy speech is dynamically mixed during the training phase, with the \xx{snr} values ranging from -5 dB to 15 dB. The trained models are first evaluated using the no-reverb test set, which consists of 150 test samples. Each file in this set contains a 10-second speech clip. The \xx{snr} of the test files in \xx{dns} test set ranges from 0 dB to 19 dB with 1 dB interval and the mean \xx{snr} of the test set is 9 dB. Additionally, to investigate the influence of \xxx{snr} on the \xx{dgru}-based models, the network is further tested on a dataset with noisy speech \xxx{snr} at \{-5, 0, 5\} dB, with 150 test utterances at each \xx{snr} level.

All training and test signals are sampled at 16 kHz. The complex spectrum for each signal is derived by applying a 320-point Fast Fourier Transform (FFT) on each frame with a size of 20\,ms and an overlap of 10\,ms.

\subsection{Evaluation Metrics}
The output of the speech enhancement algorithm is evaluated in terms of: 1) wide band PESQ~\cite{rix2001perceptual}; 2) ESTOI~\cite{jensen2016algorithm}; 3) Scale-invariant signal-to-noise ratio (SISNR) and 4) \xx{dns} OVAL~\cite{reddy2021DNSmos};. For all evaluation metrics, higher values denote better enhancement performance.

\section{Results}
\label{sec:results}

Table~\ref{tab:all_results} shows the results of speech enhancement for all tested models on the \xx{dns} test set. The results for the conventional GRU-based models (where $\mathcal{P}=100\%$) form baselines. Overall, there are only slight differences in speech enhancement across different $\mathcal{P}$ values for different models. Only when $\mathcal{P}$ is reduced to 25\% we observe a drop in PESQ scores. 

We conducted a series of hypothesis tests\footnote{Specifically employing the Mann-Whitney U test~\cite{mann1947test}, due to the non-normality of the test data.}, particularly focusing on PESQ where the most noticeable variations were observed, to determine if these variations are statistically significant. These tests compare the PESQ results at each $\mathcal{P}$-pair ($\mathcal{P}_A - \mathcal{P}_B$) within the same model. Test results (p-value) below 0.05 signifies that the differences between two tested $\mathcal{P}$ setups are statistically significant.

\begin{table}[t]
\caption{Results of hypothesis tests (see \cref{sec:results}) in terms of $p$-values for different models on each $\mathcal{P}_A-\mathcal{P}_B$ pair. Only \textbf{bold} values denote the PESQ difference of two tested models is statistically significant.}
\vspace*{-2.5mm}

\label{tab:p}
\centering
\begin{tabular}{ccccc}
\toprule
\multirow{2}{*}{Model} & \multirow{2}{*}{$\mathcal{P}_A$ (\%)} & \multicolumn{3}{c}{$\mathcal{P}_B$ (\%)} \\ \cmidrule(l){3-5} 
                       &                                & 75   & 50   & 25            \\ \midrule \midrule
\multirow{3}{*}{GRU}   & 100                          & 0.258  & 0.508  & \textbf{0.035}  \\
                       & 75                           & -      & 0.265  & \textbf{0.009}  \\
                       & 50                           & -      & -      & \textbf{0.032}  \\ \midrule
\multirow{3}{*}{MPT}   & 100                          & 0.401  & 0.377  & 0.121           \\
                       & 75                           & -      & 0.469  & 0.172           \\
                       & 50                           & -      & -      & 0.205           \\ \midrule
\multirow{3}{*}{DPRNN} & 100                          & 0.622  & 0.358  & \textbf{0.042}  \\
                       & 75                           & -      & 0.241  & \textbf{0.021}  \\
                       & 50                           & -      & -      & 0.084           \\ \bottomrule
\end{tabular}%
\vspace*{-5mm}
\end{table}

Table~\ref{tab:p} shows the detailed results of these hypothesis tests. For all models, when $\mathcal{P} \geq 50\%$, the differences between the PESQ results are not significant. At these settings, 33\% of the computation costs from the \xxx{gru} can be saved, and the overall computation costs of the \textit{GRU}, \textit{MPT}, and \textit{\xx{dprnn}} models can be reduced to 69\%, 75\%, and 71\%, respectively. 
For the \textit{GRU} and \textit{\xx{dprnn}} models, there is a significant decrease in PESQ when comparing $\mathcal{P} = 25$ to $\mathcal{P} = 100$. This suggests that the observed degradation is more likely due to the inherent limitations of the \xxx{dgru}, where updating only a limited number of neurons may lead to failures in capturing input changes. Additionally, infrequent updates of some neurons might result in inaccurate representations of the current state, introducing internal noise and worsening the speech enhancement.
In contrast, for the \textit{MPT} model, no significant differences are observed between different $\mathcal{P}$ setups. This could be attributed to the presence of more non-\xx{gru} layers in the model, which might compensate for inaccuracies in the information modeled by the \xxx{dgru}, albeit with a sacrifice of less computation savings.

Fig~\ref{fig:SNR} shows the mean PESQ improvements of each model under \{-5, 0, 5\} dB \xxx{snr}. It reveals a consistent trend observed in results on the \xx{dns} test set and further demonstrates that even under lower SNR, the dynamic GRU-based models maintain similar PESQ scores compared to the conventional GRU when $\mathcal{P}\ge50\%$.

\section{Related Works and Discussions}

Previous proposed models such as the Delta \xxx{rnn}~\cite{ neil2017delta, jelvcicova2021peakrnn, gao2022spartus} also reduce computation by allowing only inputs whose change across two steps have exceeded a threshold, to update the neurons. 
By contrast, our \xx{dgrnn} method achieves computation savings by reducing the output dimension through the proposed binary select gate that updates only the $\mathcal{P}$ percent of neurons with the largest update gate, $\boldsymbol{z}$, values. 
It would be interesting to integrate both models in a unified framework in future work.
Another model, the Phased LSTM~\cite{neil2016phased}, also allows only a subset of neurons to be updated on each step. In this model, the neuron is updated only on a particular phase of an oscillation cycle. 
Additionally, the number of updated neurons varies across steps in Phased LSTM, resulting in dynamically changing computation loads per step. 
Similarly the Skip RNN model~\cite{campos2017skip, fedorov2020tinylstms, le2022inference} is designed to skip the update of all neurons in some steps, therefore the computational load also changes across steps.
In addition, these models have been tested primarily on classification tasks, and their performance on regression tasks, such as speech enhancement, remains to be studied.

\section{Conclusion}

We propose the \xx{dgrnn} architecture for compute-efficient networks running on embedded platforms with limited resource. 
The \xx{dgrnn} reduces computes by updating only a fixed percentage of neurons per step. 
The test results show that, even with highly efficient RNN-based models and in SNR conditions down to -5 dB, updating only 50\% of the \xx{gru} neurons at each step can still achieve the same speech enhancement results. 
In the future, our aim is to investigate the applicability of these models to other modalities and tasks.

\bibliographystyle{IEEEtran}
\bibliography{mybib}

\end{document}